\documentclass[aps,twocolumn,groupedaddress]{revtex4}
\usepackage{graphicx,amsfonts,color,epsfig}
\usepackage{epstopdf}
\newcommand{\RedCG}[3]{\ensuremath{\langle#1;#2\|#3\rangle}}

\newcommand{\ket}[1]{\ensuremath{\left| #1 \right\rangle}}

\newcommand{\betb}{\begin{tabular}{p{4.0cm}p{9.0cm}}}
\newcommand{\entb}{\end{tabular}}

\begin{document}
%
%
%
%
\title{ Towards an Extended Microscopic Theory for the Upper fp-shell nuclei }
\author{K. P. Drumev}
\author{C. Bahri}
\author{J. P. Draayer}
\affiliation{Department of Physics and Astronomy, Louisiana State
University, Baton Rouge, LA 70803, USA}
\author{V. G. Gueorguiev
\footnote{On leave of absence from Institute of Nuclear Research and Nuclear
Energy, Bulgarian Academy of Sciences, Sofia 1784, Bulgaria.}}
\affiliation{Lawrence Livermore National Laboratory, Livermore, CA 94551, USA}
%
\begin{abstract}
An extended SU(3) shell model that for the first time explicitly includes unique-parity 
levels is introduced. Shell-model calculations for the isotopes of $^{64}$Ge 
and $^{68}$Se are
performed where valence nucleons beyond the N=28=Z core occupy levels of
the normal parity upper-$fp$ shell ($f_{5/2},p_{3/2},p_{1/2}$) and the
unique parity $g_{9/2}$ intruder configuration. The levels of the
upper-$fp$ shell are handled within the framework of an m-scheme
basis as well as its pseudo-SU(3) counterpart, and respectively, the
$g_{9/2}$ as a single level and as a member for the complete $gds$ shell. It is 
demonstrated that the extended SU(3) approach allows one to better probe the effects 
of deformation and to account for many key properties of the system by using a highly 
truncated model space. 
\end{abstract}
\pacs{21.60.Cs, 21.60.Fw, 21.10.Re, 27.50.+e}
\maketitle

\section{Introduction}
The nuclear shell model \cite{Meyer} has been applied successfully for the 
description of various aspects of nuclear structure, in large part because it is 
based on a minimum number of assumptions. Although direct diagonalization of the 
Hamiltonian matrix in the full Hilbert space would be desirable, the dimensionality 
of such a space is often too large to allow calculations of this type to be done. 
Recently, in order to relax this restriction dramatically, various stochastic approaches, 
for instance, the Shell-Model Monte Carlo Method \cite{MonteCarlo}, have been suggested. 
Alternatively, algebraic models using the symmetry properties of the systems under 
investigation have been developed (e.g.\cite{Elliott, JPD}).  

Intruder levels are present in heavy deformed nuclei where the strong spin-orbit 
interaction destroys the underlying harmonic oscillator symmetry of the nuclear 
mean-field potential. The role they play for the overall dynamics of the system has 
been the topic of many questions and debates \cite{Escher,Bhatt,Aberg,T(E2)_def}. Until now, the problem 
has been either approached within the framework of a truncation-free toy model \cite{Escher} 
or by just considering the role of the single intruder level detached from its like-parity 
partners \cite{Bhatt,T(E2)_def}. It was argued in \cite{Escher} that particles in these levels contribute 
in a complementary way to building  the collectivity in nuclei. However, some mean-field theories
suggest that these particles play the dominant role in inducing deformation \cite{Aberg}. 
In order to build a complete shell-model theory, these levels need to be included in the 
model space especially if experimentally observed high-spin or opposite-parity states 
are to be described. 

Until recently, SU(3) shell-model calculations - real SU(3)
\cite{Elliott} for light nuclei and pseudo-SU(3) \cite{JPD} for heavy nuclei -
have been performed in either only one space (protons and neutrons filling
the same shell, e.g. the $ds$ shell) or two spaces (protons and neutrons
filling different shells, e.g. for rare-earth and actinide nuclei). Various
results for low-energy features, like energy spectra and electromagnetic
transition strengths, have been published over the years
\cite{Vargas_even, Vargas_odd, SU(3)_applications}. These applications confirm
that the SU(3) model works well for light nuclei and the pseudo-SU(3) scheme,
under an appropriate set of assumptions, for rare-earth and actinide species.
Up to now, SU(3)-based methodologies have not been applied to nuclei with mass
numbers $A=56$ to $A=100$, which is an intermediate region where conventional
wisdom suggests the break down of the  assumptions that underpin their
use in the other domains. In particular, the $g_{9/2}$ intruder
level that penetrates down from the shell above due to the strong
spin-orbit splitting appears to be as spectroscopically relevant to the overall
dynamics as the normal-parity $f_{5/2},p_{3/2},p_{1/2}$ levels.
Specifically, in this region the effect of the intruder level cannot be
ignored or mimicked through a ``renormalization'' of the normal-parity
dynamics which is how it has been handled to date. 
\begin{figure*}[t]
\includegraphics[width=0.99\textwidth]{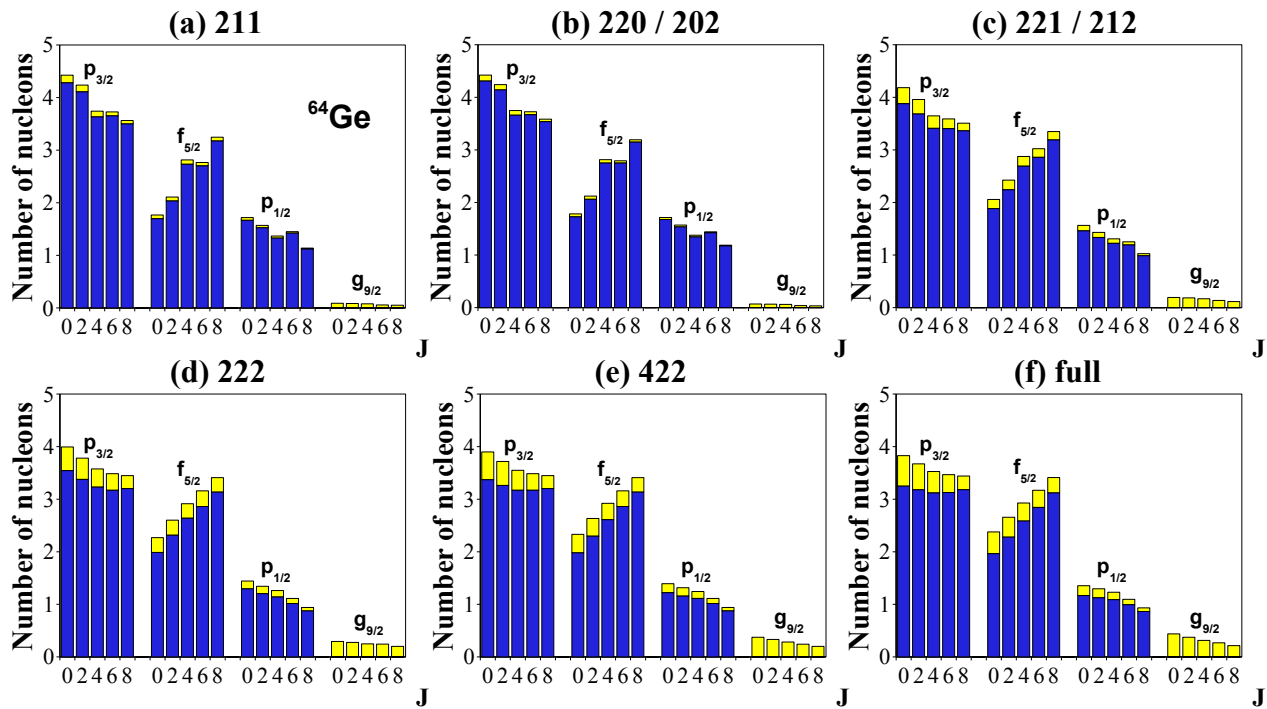}
\caption{(Color online) Single-particle occupation numbers for eigenstates of the g.s. band of
$^{64}$Ge calculated in different restricted model spaces (from (a) to (e)) and the full space (f).
The labels TPN over each restricted-space calculation represent the maximum total number of particles
T, and the maximum number of protons P (and neutrons N) allowed in the intruder $g_{9/2}$ level.}
\label{Figure_occupancies}
\end{figure*}

The upper-$fp+g_{9/2}$ shell is the lightest region of nuclei where
the intruder level must be taken into account. Its presence poses a
significant challenge and opens a sequence of questions, many of them 
still unanswered.  For example, if the pseudo-SU(3) symmetry proves
to be a good scheme for characterizing upper-$fp$ shell configurations,
should one integrate the $g_{9/2}$ intruder into this picture by
treating it as a single $j$-shell that is independent of couplings to
the other members ($g_{7/2},d_{5/2},d_{3/2},s_{1/2}$) of the $gds$
shell, or should one take the complete $gds$ shell into account? While 
the treatment of the $g_{9/2}$ intruder as a single additional orbit 
is possible for only a few exceptional nuclei within the m-scheme shell-model 
calculations, an SU(3) approach to the problem could allow the inclusion of the 
$g_{9/2}$ intruder as a member of the full gds shell and can be applied 
to all upper-fp nuclei.
It is the purpose of this work to introduce and establish the benefits of a new and extended
SU(3) shell model which, for the first time, explicitly includes particles
from the complete unique-parity sector and therefore can be used to explore the role that
intruder levels play in the dynamics of the system.
Calculations are performed for two nuclei which are of major importance in astrophysics,
namely, the waiting-point $N=Z$ nuclei $^{64}$Ge and $^{68}$Se \cite{Schatz}.
In addition, $^{68}$Se is known to be among the nuclei for which shape coexistence
effects have been reported \cite{Kaneko_68Se, Sun}. Both the strengths and the limitations of the model
are demonstrated and discussed.

First, a motivation for our approach is presented and the 
appropriate choice of model space is established by calculating, using a realistic
Hamiltonian, the occupancies of the single-particle orbitals and the symmetry 
properties of these typical upper-fp shell nuclei. Then, following a brief introduction of the 
basics of the extended SU(3) model, calculations for the energy spectra, B(E2) transition strengths 
and the wave function content are performed and results are compared with the realistic predictions. 
Finally, discussion about various future applications of the model is included.  

\section{A reasonable approach for the description of upper fp-shell nuclei}
To benchmark the benefit of the SU(3) scheme in this region (pseudo-SU(3) 
for the upper-$fp$ shell and normal SU(3) for the $g_{9/2}$ configurations 
extended to the full $gds$ shell), we first generated results in a standard 
m-scheme representation for both nuclei, $^{64}$Ge and $^{68}$Se, with the 8 
(4 protons + 4 neutrons) and 12 (6 protons + 6 neutrons) valence nucleons, 
respectively, distributed across the $p_{1/2},p_{3/2},f_{5/2},g_{9/2}$ 
model space with the $f_{7/2}$ level considered to be fully occupied and part of a $^{56}$Ni 
core. The Hamiltonian we used is a G-matrix with a phenomenologically adjusted monopole
part \cite{Caur, Abzouzi} that in many cases describes the experimental energies 
reasonably well. Specifically, this 
upper-$fp+g_{9/2}$ shell interaction was succesfully used in the past to obtain quite good 
results for nuclei like $^{62}$Ga \cite{Vincent}, and $^{76}$Ge and $^{82}$Se \cite{Caur}. 
Later, it was applied for exploring the pseudo-SU(4) symmetry in the region from 
the beginning of the upper fp-shell up to $N=30$ and for describing beta decays 
\cite{Van Isacker}.   
\begin{figure}[htb]
\includegraphics[width=0.45\textwidth]{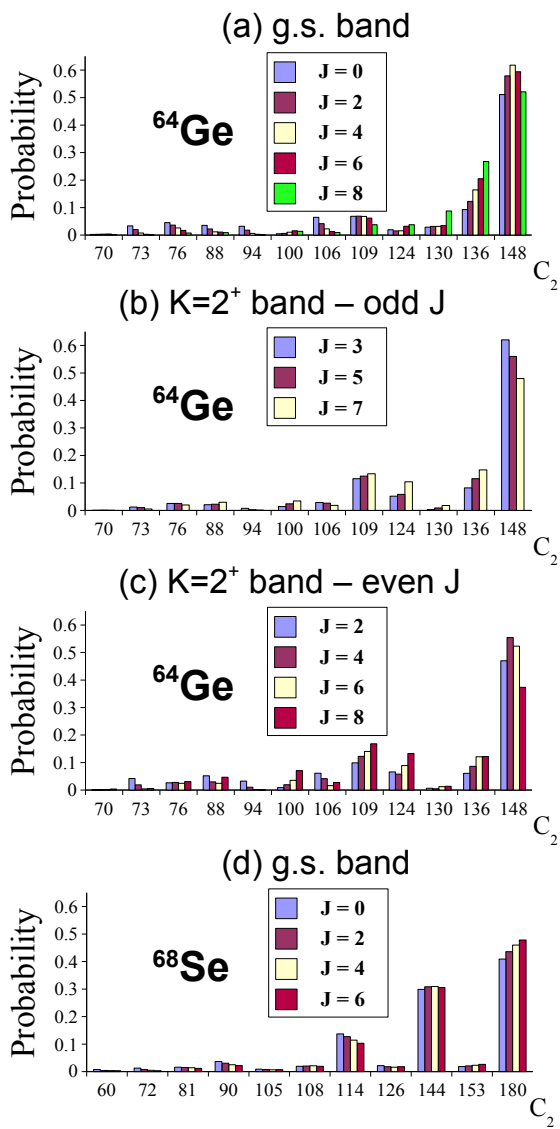}
\caption{(Color online) Pseudo-SU(3) content of the low-lying states in (a) the g.s. band of $^{64}$Ge, 
 (b) and (c) - the $K=2^{+}$ band of $^{64}$Ge, and (d) the g.s. band of $^{68}$Se using the renormalized
counterpart of the G-matrix realistic interaction.}
\label{Figure_symmetry}
\end{figure}

Calculations with different cuts of the full model space were done in 
order to estimate the occupancy of the single-particle levels and thus to 
evaluate the relative importance of various configurations for describing essential 
nuclear characteristics. In Fig. \ref{Figure_occupancies}, a comparison is made between 
the occupancies for $^{64}$Ge as determined in a restricted basis, where at most two 
particles (protons/neutrons) are allowed in the $g_{9/2}$ level and the 
full-space resuls. The upper (yellow) bars show the  contribution to occupations 
from basis states with an occupied intruder level while the lower (blue)
portion represents those where the intruder level is empty. The calculated results 
suggest that  the occupancy probability for the intruder level is approximately 0.3 particles for 
the low-lying states of $^{64}$Ge.  
Calculations with no particles allowed in the intruder level or with just one
identical-particle (or proton-neutron) pair (Fig. \ref{Figure_occupancies} (a) and (b))
cannot describe either its occupancy or the gradual change in the occupancy of the 
single-particle levels in the ground-state (g.s.) band that is found in the
full-model-space results. However, using a restricted space with at most two
identical particles occupying the intruder level 
(in Fig. \ref{Figure_occupancies}(e)) is sufficient to describe both
features as well as the low-energy spectrum and the B(E2) transition
strengths \cite{Drumev}.  
Similar results were observed for the $K=2^{+}$ band of this nucleus. 
As expected, calculations for $^{68}$Se performed in a truncated basis
with at most 2 nucleons allowed in the intruder level produce a slighlty
higher value of the $g_{9/2}$ occupancy compared to $^{64}$Ge.

Next, the goodness of the pseudo-SU(3) symmetry in these nuclei was tested 
using a renormalized version of the same realistic interaction in the $pf_{5/2}$ space \cite{Van Isacker}. 
The matrix of the second-order Casimir  
operator of pseudo-SU(3), $C_{2} = \frac{1}{4}(3 \hat{L}^{2}+Q.Q)$, was generated and the method 
of moments \cite{Whitehead} used to diagonalize this matrix by starting the Lanzcos 
procedure with specific eigenvectors of the Hamiltonian for which a pseudo-SU(3) decomposition 
was desired. Although the procedure provides distributions only over the $C_{2}$ values (and not 
over the actual $(\lambda,\mu)$ irreducible representations (irreps)) this analysis is quite useful and
gives valuable information about the structure of certain eigenstates.

The distribution of the second order Casimir operator $C_2$ of pseudo-SU(3) yields 
contributions of about 50-60\% from the leading pseudo-SU(3) irrep in the g.s. band
of $^{64}$Ge (Fig. \ref{Figure_symmetry}(a)) which suggests that the pseudo-SU(3)
symmetry is quite good. In the $K=2^{+}$ band (Fig. \ref{Figure_symmetry} (b) and (c))
this contribution appears to be somewhat lower, ranging from approximately 37\% for
the $8_{2}^{+}$ state to about 62\% in the $3_{1}^{+}$ state. The analysis also
reveals that using only five irreps which have the highest $C_{2}$ value one may take
account of at least 70\% and up to about 95\% of the wavefunction for the states in
these bands. 

In the case of $^{68}$Se, the outcome turns out to be quite similar for the states
from the g.s. band (Fig. \ref{Figure_symmetry}(d)). Although the irreps with the
maximal value of $C_{2}=180$ participate with only between about 40\% and 50\%, the
first eleven irreps with distinct values of $\lambda$ and $\mu$ account for 88-93\% of
the wavefunction. In addition, the $0_{3}^{+}$ state at 2.51 MeV is also dominated
($64\%$) by irreps with the biggest $C_{2}$ value. However, other states are predicted
to be highly-mixed SU(3) configurations. This includes the $0_{2}^{+}$ state found at
1.05 MeV - a value very similar to the ones reported in \cite{Kaneko_68Se, Sun} for a
low-lying state of prolate shape. A recent analysis reveales that many low-lying
states in other $N\sim Z$ nuclei also have good pseudo-SU(3) symmetry
which further underscores the value of using symmetry-based truncation
schemes \cite{Drumev}.

In summary, the outcome of the m-scheme study has demonsrtated that only part of the
configurations are relevant for the structure of the low-lying states in the upper-fp
shell nuclei. This along with the pseudo-SU(3) spectral decomposition of the states
establishes the validity of a SU(3)-based truncation scheme in $N \sim Z$ upper-fp
shell nuclei. As for the unique-parity spaces, the so-called quasi-SU(3) concept
\cite{Vargas_even,Vargas_odd,Quasi_SU(3)} may also be applicable, however, this question is deferred to
a follow-on investigation. Since (as we will see below) the role of the unique-parity
spaces for the nuclei we deal with, is to introduce only some high-order effects, it
is relatively safe to accept this to be true throughout the current work.    
\section{An Extended SU(3) Model with Explicitly Included Intruder Levels}
 Following the series of arguments and motivations presented in the previous section, we can
now introduce the basics of the extended SU(3) shell model. Like its early precursors
\cite{Elliott, JPD}, it is also a microscopic theory in the  sense that both SU(3) generators
- the angular momentum ($L_{\mu},\mu=0, \pm1$) and quadrupole ($Q_{\mu};\mu=0, \pm1, \pm2$)
operators - are given in terms of individual nucleon coordinate and momentum variables.
However, the model space has a more complicated structure than the one used in earlier models
based on the SU(3) symmetry. Specifically, it consists of two parts for each particle type,
a normal (N) parity pseudo-shell ($f_{5/2}, p_{3/2}, p_{1/2}$ $\rightarrow$
$\tilde{d}_{5/2},\tilde{d}_{3/2},\tilde{s}_{1/2}$) and a unique (or abnormal) (U) parity
shell composed of all levels of opposite parity from the $gds$ shell above. 
(Since the normal-unique space distinction will be obvious from context, we will not place 
tildes over pseudo-space labels as is normally done.) 

 The many-particle basis states 
\begin{equation} 
\ket{\{a_\pi;a_\nu\} \rho (\lambda,\mu) \kappa L, \{S_{\pi},S_{\nu}\}S;JM}
\end{equation}
\noindent are built as SU(3) proton ($\pi$) and neutron ($\nu$) coupled configurations with 
well-defined particle number and good total angular momentum. Here, the proton and neutron 
quantum numbers are indicated by
$a_\sigma=\{a_{\sigma N}, a_{\sigma U}\}\rho_\sigma
(\lambda_\sigma,\mu_\sigma)$, where the $a_{\sigma\tau}=
N_{\sigma \tau} [f_{\sigma \tau} ] \alpha_{\sigma \tau}
(\lambda_{\sigma \tau}, \mu_{\sigma \tau})$ are
the basis-state labels for the four spaces in the model ($\sigma$
stands for $\pi$ or $\nu$, and $\tau$ stands for N or U). In the last expression, 
$N_{\sigma \tau}$ denotes the number of particles in the corresponding space, 
[$f_{\sigma \tau}$] - the spatial symmetry label and $(\lambda_{\sigma \tau},\mu_{\sigma \tau})$ -
the SU(3) irrep label. 
Multiplicity indices $\alpha_{\sigma \tau}$ and $\rho_{\sigma}$ count different occurences
of $(\lambda_{\sigma \tau},\mu_{\sigma\tau})$ in [$f_{\sigma \tau}$] and in the product
$\{(\lambda_{\sigma N},\mu_{\sigma N}) \times
(\lambda_{\sigma U},\mu_{\sigma U})\}\rightarrow (\lambda_{\sigma}, \mu_{\sigma})$,
respectively. First, the particles from the normal and
the unique spaces are coupled  for both protons and neutrons. Then, the resulting
proton and neutron irreps are coupled to a total final set of irreps. The total
angular momentum $J$ results from the coupling of the total orbital angular
momentum $L$ with the total spin $S$. The $\rho$ and $\kappa$ are, respectively,
the multiplicity indices for the different occurences of
($\lambda, \mu$) in $\{(\lambda_{\pi},\mu_{\pi}) \times 
(\lambda_{\nu}, \mu_{\nu})\}$ and $L$ in ($\lambda,\mu$).

The Hamiltonian
\begin{eqnarray} H &=& \sum_{\sigma,\tau}(H_{sp}^{\sigma
\tau}-GS^{\sigma \tau \dag}S^{\sigma
\tau})-\frac{\chi}{2}:Q.Q:+aJ^{2}+bK_{J}^2 \nonumber \\
&-& G (\sum_{\sigma , \tau \neq \tau^{\prime}}S^{\sigma \tau
\dag}S^{\sigma \tau^{\prime}}
+\sum_{\tau , \tau^{\prime}}S^{\pi \nu , \tau
\dag}S^{\pi \nu , \tau^{\prime}}) 
\label{Hamiltonian}
\end{eqnarray}
\noindent includes spherical Nilsson single-particle energies 
\begin{equation}
H_{sp}^{\sigma
\tau} = \sum_{i_{\sigma \tau}}(H_{0}+C_{\sigma \tau}{\bf l}_{i_{\sigma \tau}}.
{\bf s}_{i_{\sigma \tau}}+D_{\sigma \tau}{\bf l}_{i_{\sigma \tau}}^{2})
\end{equation}
\noindent as well as the quadrupole-quadrupole and pairing interactions (within a shell
and between shells) plus
two rotor-like terms that are diagonal in the SU(3) basis. In general, the harmonic
oscillator term, $H_{0}=\hbar \omega (\eta_{i_{\sigma \tau}}+ \frac{3}{2})$ where
$\hbar \omega \approx \frac{41}{A^{1/3}}$ \cite{Ring},
is essential and its contribution does not cancel out when more than one
possible distribution of particles  over the shells is involved.
The colons in the quadrupole operator notation represent normal-ordered operator since
all the one-body effects have already been taken into account by the
single-particle terms in the Hamiltonian. In addition, in first approximation,
the quadrupole operator in the normal-parity spaces is related to its pseudo counterpart
by $Q_{\sigma N}\approx\frac{\tilde{\eta}+1}{\tilde{\eta}}\tilde{Q}_{\sigma N}$ with $\tilde{\eta}$
equal to 2 for both protons and neutrons and $Q=Q_{\pi N}+Q_{\pi U}+Q_{\nu N}+Q_{\nu U}
\approx 1.5\;\tilde{Q}_{\pi N}+Q_{\pi U}+1.5\;\tilde{Q}_{\nu N}+Q_{\nu U} $.
\begin{table}[h]
\caption{ Parameters (in MeV) used in the extended SU(3) model Hamiltonian.
\label{TABLE_parameters}}
\begin{ruledtabular}
\begin{tabular}{ccccc}
 Nucleus & $G$ & $\chi$ & $a$ & $b$ \\
\hline
  $^{64}$Ge & $0.280$ & $0.0176$ & $-0.002$ & $0.020$ \\
  $^{68}$Se & $0.263$ & $0.0152$ & $-0.002$ & $0.000$
\end{tabular}
\end{ruledtabular}
\end{table}
\begin{table*}[th]
\caption{ The irreps in the coupled proton-neutron model space for $^{64}$Ge
and $^{68}$Se that were used in the extended SU(3) shell-model calculations.
The subscripts for each spin value denote the multiplicity, that is, the number
of different ways the corresponding irrep can be constructed.
\label{TABLE_irreps}}
\begin{ruledtabular}
\begin{tabular}{clllll}
$[N_{\pi N}, N_{\pi U}; N_{\nu N}, N_{\nu U}]$ &
\multicolumn{5}{c}{total$\;(\lambda,\mu)S_{\textrm{multiplicity}}$}\\
\hline
\multicolumn{6}{c}{$^{64}$Ge} \\
   $[4,0;4,0]$ & $(8,4)0_{1}$ & $(9,2)0_{1},1_{2}$ & $(10,0)0_{2},1_{1},2_{1}$ 
                 & $(6,5)0_{1},1_{2}$ & $(7,3)0_{6},1_{8},2_{2}$ \\
   & & & & & \\
   $[4,0;2,2]$ & $(16,2)0_{2}$ & $(17,0)1_{2}$ & $(14,3)0_{6},1_{6}$ 
                 & $(15,1)0_{12},1_{15},2_{4}$ & $(12,4)0_{18},1_{18},2_{4}$ \\
   $([2,2;4,0])$ & & & & & \\
   & & & & & \\
   $[3,1;3,1]$ & $(16,2)0_{2},1_{3},2_{1}$ & $(17,0)0_{2},1_{3},2_{1}$ & $(14,3)0_{8},1_{15},2_{5}$ 
                 & $(15,1)0_{16},1_{27},2_{10}$ & $(12,4)0_{28},1_{43},2_{15}$ \\
   & & & & & \\
   $[2,2;2,2]$ & $(24,0)0_{1}$ & $(22,1)0_{3},1_{4}$ & $(20,2)0_{16},1_{18},2_{4}$ 
                 & $(21,0)0_{9},1_{18},2_{4}$ & $(18,3)0_{42},1_{64},2_{16}$ \\
   $$ & $$ & $$ & $$ & $$ & $$ \\
\hline
\multicolumn{6}{c}{$^{68}$Se} \\
   $[6,0;6,0]$ & $(12,0)0_{1}$ & $(0,12)0_{1}$ & $(9,3)0_{2},1_{2}$ 
                 & $(3,9)0_{2},1_{2}$ & $(10,1)0_{1},1_{2}$ \\ 
                 & $(1,10)0_{1},1_{2}$ & $(6,6)0_{4},1_{3},2_{1}$ & $(7,4)0_{6},1_{11},2_{3}$ 
                 & $(4,7)0_{6},1_{11},2_{3}$ & $(8,2)0_{8},1_{12},2_{4}$ \\ 
                 & $(2,8)0_{8},1_{12},2_{4}$ & $$ & $$ & $$ & $$ \\ 
   & & & & & \\
   $[6,0;4,2]$ & $(18,2)0_{2}$ & $(19,0),1_{2}$ & $(15,5)0_{2},1_{2}$ 
                 & $(16,3)0_{8},1_{12},2_{2}$ & $(12,8)0_{2}$ \\
   $([4,2;6,0])$ & $(17,1)0_{12},1_{16},2_{4}$ & $(13,6)0_{10},1_{16},2_{4}$ & & & \\ 
   & & & & & \\
   $[5,1;5,1]$ & $(18,2)0_{2},1_{3},2_{1}$ & $(19,0)0_{2},1_{3},2_{1}$ & $(15,5)0_{4},1_{6},2_{2}$ 
                 & $(16,3)0_{16},1_{27},2_{13}$ & $(12,8)0_{2},1_{3},2_{1}$ \\
                 & $(17,1)0_{18},1_{30},2_{14}$ & $(13,6)0_{20},1_{33},2_{15}$ & & & \\
   & & & & & \\
   $[4,2;4,2]$ & $(24,4)0_{1}$ & $(25,2)0_{1},1_{2}$ & $(26,0)0_{2},1_{1},2_{1}$ 
                 & $(22,5)0_{3},1_{4}$ & $(23,3)0_{14},1_{20},2_{6}$ \\
                 & $(24,1)0_{19},1_{31},2_{13}$ & $(20,6)0_{16},1_{18},2_{6}$ & $$ & $$ & $$ \\
\end{tabular}
\end{ruledtabular}
\end{table*} 

The second line in Eq.(\ref{Hamiltonian}) consists of pairing terms that are 
included for the first time in SU(3) shell-model calculations. In particular, the first term 
represents the scattering of an identical-particle pair between the normal- and unique-parity spaces. 
The second one stands for the proton-neutron pairing (or simply pn-pairing) interaction within the 
normal- or unique-parity space 
(terms with $\tau = \tau^{\prime}$) and for the pn-pair scattering between the normal- and 
unique-parity spaces (terms with $\tau \neq \tau^{\prime}$). The formulae used in this work for each type of pairing operator can be found in the Appendix \ref{appendix:pairing_terms}.
It is worth mentioning that, contrary to many calculations 
involving the pairing interaction, these expressions are exact. Finally, the two rotor-like terms
$J^2$ and $K_{J}^{2}$ (the square of the total angular momentum and its projection on the intrinsic
body-fixed axis) are used to ``fine tune" the energy spectra, adjusting  the moment of inertia of 
the g.s. band and the position of the $K=2^{+}$ bandhead, respectively. Their strengths are the only
two parameters fitted in this work.

The single-particle terms together with the proton, neutron and proton-neutron pairing interactions mix the 
SU(3) basis states, which allows for a realistic description of the energy spectra of the 
nuclei. The values of the parameters used in Hamiltonian (\ref{Hamiltonian}) can be found in Table \ref{TABLE_parameters}.
The single-particle energies in the  Hamiltonian for the normal spaces are fixed  with the numbers provided
by the upper-fp shell single-particle energies and for the strengths in the unique-parity spaces the numbers
from systematics are used \cite{Ring}. The values for the parameters $G$ and $\chi$ in the Hamiltonian which are taken from
\cite{Kaneko_64Ge} are found to be in agreement with  the ones \cite{Ring, Zuker} used in previous calculations 
for some ds-shell and rare-earth nuclei. For simplicity, we take both identical-particle and proton-neutron 
pairing strengths to be equal.  
\section{Results and Discussion}
Calculations within the framework of the extended SU(3) model were performed
using irreps from 5 types of configurations - for example,
$[N_{\pi N}, N_{\pi U};N_{\nu N}, N_{\nu U}]$ = $[4,0;4,0]$, $[4,0;2,2]$, 
$[2,2;4,0]$, $[3,1;3,1]$ and $[2,2;2,2]$ for the $^{64}$Ge case. (See Table
\ref{TABLE_irreps} where the list of configurations for $^{68}$Se is also
given.) For each of these groups, irreps in the proton and neutron spaces
with (pseudo-) spin $S_{\sigma \tau}=0,~1/2,~1$ and $3/2$ in both the
normal- and the unique-parity spaces were generated. Then, from all the
possible couplings between these we chose those with the largest value of
the second order Casimir operator of SU(3) and spin $S=0$, $1$ and $2$.
Here, we present results obtained with five (seven) coupled proton-neutron
irreps  with distinct values of $\lambda$ and $\mu$ for each distribution
of particles between the normal and unique spaces for $^{64}$Ge ($^{68}$Se).
(This number was even pushed up to eleven for the [6,0;6,0] configuration
in $^{68}$Se). The complete set, listed in Table \ref{TABLE_irreps}, consists
of $492\;(580)$ coupled irreps in the case of $^{64}$Ge ($^{68}$Se).
This  number is bigger by a factor of about 20 than the one typically handled
up to now within the framework of the SU(3) model. Some of the coupled irreps
can be constructed in more than one way. For example, the irrep
$(\lambda,\mu)S=(10,0)0$ can be obtained by coupling the 
$(\lambda_{\pi},\mu_{\pi})S_{\pi}\times (\lambda_{\nu},\mu_{\nu})S_{\nu}=(4,2)0
\times (4,2)0$ or $(5,0)1 \times (5,0)1$ proton and neutron irreps.
\begin{figure}[h]
\includegraphics[width=0.45\textwidth]{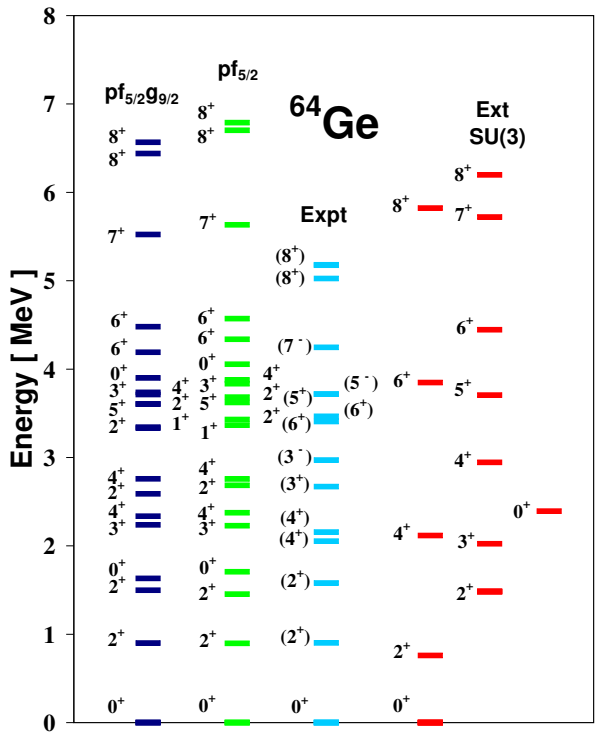}
\caption{(Color online) Low-energy spectra of $^{64}$Ge obtained with (from left to right) the 
realistic interaction in the full $pf_{5/2}g_{9/2}$ and $pf_{5/2}$ model spaces compared with experiment 
\cite{NNDC} and the extended-SU(3)-model results.}
\label{Figure_spectrum_64Ge}
\end{figure}
\begin{figure}[b]
\includegraphics[width=0.45\textwidth]{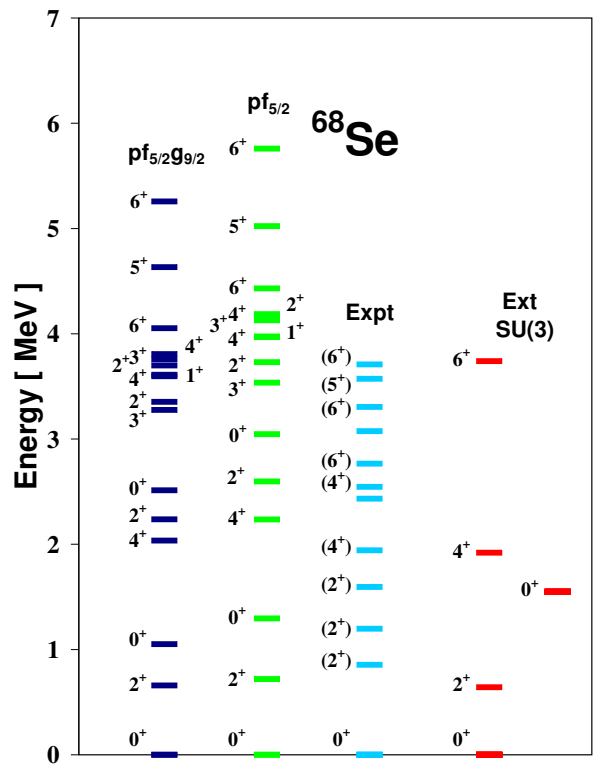}
\caption{(Color online) Low-energy spectra of $^{68}$Se obtained with (from left to right) the 
realistic interaction in the restricted $pf_{5/2}g_{9/2}$ (at most 2 particles allowed in the intruder 
$g_{9/2}$ level) and full $pf_{5/2}$ model spaces compared
with experiment \cite{NNDC} and the extended-SU(3)-model results.}
\label{Figure_spectrum_68Se} 
\end{figure}

For both $^{64}$Ge and $^{68}$Se, proton-neutron configurations with no particles in the
unique space are found, as expected, to lie lowest and determine, by-and-large, the structure
of the low-lying eigenstates. Only a small portion of all proton-neutron coupled irreps -
27 (112) in the case of $^{64}$Ge ($^{68}$Se) - belong to these types of
configurations, which we will refer to as the dominant ones. Since the only possible irrep in
the unique-parity spaces for this case is $(\lambda_{\pi U},\mu_{\pi U})=(0,0)$
(and $(\lambda_{\nu U},\mu_{\nu U})=(0,0)$), these configurations are the exact pseudo analog
of the ones encountered in the ds-shell nuclei $^{24}$Mg and $^{28}$Si which have been studied
earlier \cite{Vargas_even}. Using only the principal part of the Hamiltonian (\ref{Hamiltonian}),
namely, the part with both rotor term strengths equal to zero, we are able to provide a good
description of  the low-lying states. Specifically, all the energies from the g.s. bands (with
the exception of the $2_{1}^{+}$ state in $^{68}$Se) differ by no more than 15\% from the
experimental values \cite{NNDC}. In order to conform with this result and prevent any further
changes in the structure of the wave function, the range of values for the parameters $a$ and $b$ were
severely restricted so that these terms  only introduce small  (``fine tuning'') changes to the overall fit.

\begin{figure}
\includegraphics[width=0.45\textwidth]{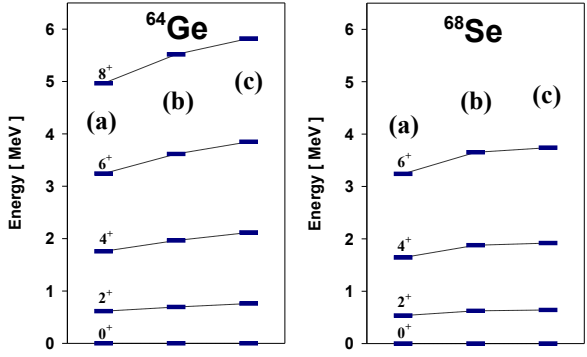} 
\caption{(Color online) Role of the pn-pairing and the pair-scattering terms
for the states in the g.s. band of $^{64}$Ge and $^{68}$Se: (a) both pair-scattering
and pn-pairing contributions excluded, (b) only pair-scattering 
contribution excluded and (c) total interaction.}
\label{Figure_energies_pairing} 
\end{figure}

Proton-neutron configurations with two and four particles in the unique-parity space
prevail at higher energies. The former starts to dominate from about 3.5 MeV (5 MeV)
in the case of $^{64}$Ge ($^{68}$Se), usually for states of higher spin values,
and the latter at even higher energies. The amount of mixing found between
configurations with different distribution
of particles is due to the pair-scattering interactions between the normal- and unique-parity spaces.
While the expected behavior of the unique-parity-space occupancy is observed - it goes up with the rise
of both $G$ and/or $\chi$ strengths - with the choice of parameters from Table \ref{TABLE_parameters} the
absolute values appear to be underestimated by at least a factor of 3 in $^{64}$Ge (see the result shown
in Fig. \ref{Figure_occupancies}) and by even more for the case of $^{68}$Se. This may indicate that the
model space has to be further expanded to acommodate more mixing from the pair-scattering interaction
terms, that the pn-interaction indeed should enter with different, possibly bigger strength than the
identical-particle pairing, or the possible need to include other terms in the Hamiltonian like the
quadrupole and isoscalar pairing interactions.

Results for the excitation spectra of $^{64}$Ge are presented in Fig. \ref{Figure_spectrum_64Ge}.
The realistic G-matrix interaction gives a reasonable result for the low-lying states consistent
with the one obtained in \cite{Kaneko_68Se, Kaneko_64Ge}. Moreover, a description of a similar
quality is provided by the extended SU(3) model. The existence of two prolate bands, as predicted
by the calculations with the realistic interactions, is also observed, that is, a g.s. $K=0^{+}$ and an
excited $K=2^{+}$ band, both dominated by the $(8,4)$ irrep. The first excited
$0^{+}$ ($0_{2}^{+}$) state, not reported yet experimentally, is found at 2.39 MeV which
is higher than the prediction made by the realistic interactions.

  Consistent with the outcome for $^{64}$Ge, in the case of $^{68}$Se we found a
reasonable description for the energies of the states from the g.s. band
(Fig. \ref{Figure_spectrum_68Se}). Even the use of a restricted space with at most
2 nucleons allowed in the intruder $g_{9/2}$ level produces result which reflects
some basic characterisics of the full-space spectrum reported in \cite{Kaneko_68Se}.
For example, the first excited $0^{+}$ state ($0_{2}^{+}$) is also positioned below
the $2_{2}^{+}$ state. A new feature observed in our results is that the $0_{3}^{+}$
state at 2.51 MeV was found to be dominated by the shapes with $C_{2}=180$ (see
Section II). Within the framework of the extended SU(3) shell model, $^{68}$Se is
predicted to be a mid-shell nucleus, a fact which may explain the existence of
shape coexistence effects. Unlike the case of $^{28}$Si \cite{Vargas_even}, now the
g.s. band is dominated by the irrep $(12,0)$ which corresponds to a prolate shape.
This result mainly follows due to the presence of the orbit-orbit terms in the
Hamiltonian and is in agreement with some earlier discussions \cite{Sahu, Fischer}.
Specifically, it favors the scenario in which the lower eigenstates in the g.s.
band are prolate and throughout the band the shape changes to oblate \cite{Sahu}.
Because of the nature of the leading representation, the model can not easily account
for a $K=2^{+}$ band with the same shape characteristic, neither can it give a simple
explanation for a low-lying $K=0^{+}$ band, facts which are in support of the
realistic prediction made in Section II for a highly-mixed nature of the $0_{2}^{+}$
as well as many other low-lying states in this nucleus. With only the $J^{2}$ term
used in the Hamiltonian for adjusting the energies, the $0_{2}^{+}$ state is predicted
by the model at 1.55 MeV.

\begin{table}[h]
\caption{B(E2) transition strengths for $^{64}$Ge in units of $e^{2}fm^{4}$ calculated using 
the G-matrix interaction in full $pf_{5/2}$ and $pf_{5/2}g_{9/2}$ model spaces, and the extended
SU(3) model.  Entries in parentheses show the result when only the normal spaces are used in the 
calculations.
\label{TABLE_strengths_64Ge}}
\begin{ruledtabular}
\begin{tabular}{cccc}
 $(J+2)^{+}\rightarrow J^{+}$ & $pf_{5/2}$ & $pf_{5/2}g_{9/2}$ & Ext. SU(3) \\
\hline
  $2_{g.s.}^{+}\rightarrow0_{g.s.}^{+}$  & $257.22$ & $253.91$ & $292.80~(280.10)$ \\
  $4_{g.s.}^{+}\rightarrow2_{g.s.}^{+}$  & $332.54$ & $342.51$ & $346.26~(334.10)$ \\
  $6_{g.s.}^{+}\rightarrow4_{g.s.}^{+}$  & $340.51$ & $356.92$ & $380.39~(370.56)$ \\
  $8_{g.s.}^{+}\rightarrow6_{g.s.}^{+}$  & $303.31$ & $320.14$ & $273.84~(268.08)$ \\
  $$  & $$ & $$ & $$ \\
  $4_{\gamma}^{+}\rightarrow2_{\gamma}^{+}$  & $89.26$ & $93.13$ & $67.25~(65.73)$ \\
  $6_{\gamma}^{+}\rightarrow4_{\gamma}^{+}$  & $164.23$ & $144.19$ & $207.18~(204.78)$ \\
  $8_{\gamma}^{+}\rightarrow6_{\gamma}^{+}$  & $92.12$ & $84.38$ & $74.79~(79.39)$ \\
\hline
$(J+1)^{+}\rightarrow J^{+}$  &  &  & \\
\hline
  $3_{\gamma}^{+}\rightarrow2_{\gamma}^{+}$  & $371.15$ & $357.79$ & $505.27~(493.39)$ \\
  $5_{\gamma}^{+}\rightarrow4_{\gamma}^{+}$  & $238.48$ & $240.40$ & $137.48~(135.48)$ \\
  $7_{\gamma}^{+}\rightarrow6_{\gamma}^{+}$  & $159.44$ & $161.24$ & $10.26~(10.17)$ \\
\hline
$J_{\alpha}^{+}\rightarrow J_{\beta}^{+}$ & & & \\
\hline
  $2_{\gamma}^{+}\rightarrow0_{g.s.}^{+}$  & $1.98$ & $1.42$ & $5.71~(5.74)$ \\
  $2_{\gamma}^{+}\rightarrow2_{g.s.}^{+}$  & $251.68$ & $241.41$ & $183.16~(178.96)$ \\
  $3_{\gamma}^{+}\rightarrow2_{g.s.}^{+}$  & $4.21$ & $3.40$ & $9.90~(9.93)$ \\
  $4_{\gamma}^{+}\rightarrow4_{g.s.}^{+}$  & $72.10$ & $74.69$ & $47.11~(46.89)$ \\
  $4_{\gamma}^{+}\rightarrow2_{g.s.}^{+}$  & $18.86$ & $19.31$ & $6.70~(6.75)$ \\  
\end{tabular}
\end{ruledtabular}
\end{table}
\begin{table}[h]
\caption{B(E2) transition strengths for the states in the g.s. band of $^{68}$Se in
units of $e^{2}fm^{4}$ calculated using the G-matrix interaction in full
$pf_{5/2}$ model space and the extended SU(3) model. Entries in parentheses
show the result when only the normal spaces are used in the calculations.
\label{TABLE_strengths_68Se}}
\begin{ruledtabular}
\begin{tabular}{ccc}
 $(J+2)^{+}\rightarrow J^{+}$ & $pf_{5/2}$ & Ext. SU(3) \\
\hline
  $2_{g.s.}^{+}\rightarrow0_{g.s.}^{+}$  & $322.71$ & $354.17~(346.37)$ \\
  $4_{g.s.}^{+}\rightarrow2_{g.s.}^{+}$  & $448.07$ & $486.65~(477.18)$ \\
  $6_{g.s.}^{+}\rightarrow4_{g.s.}^{+}$  & $441.58$ & $473.89~(467.09)$ \\ 
\end{tabular}
\end{ruledtabular}
\end{table}
%
%
\begin{figure}
\includegraphics[width=0.45\textwidth]{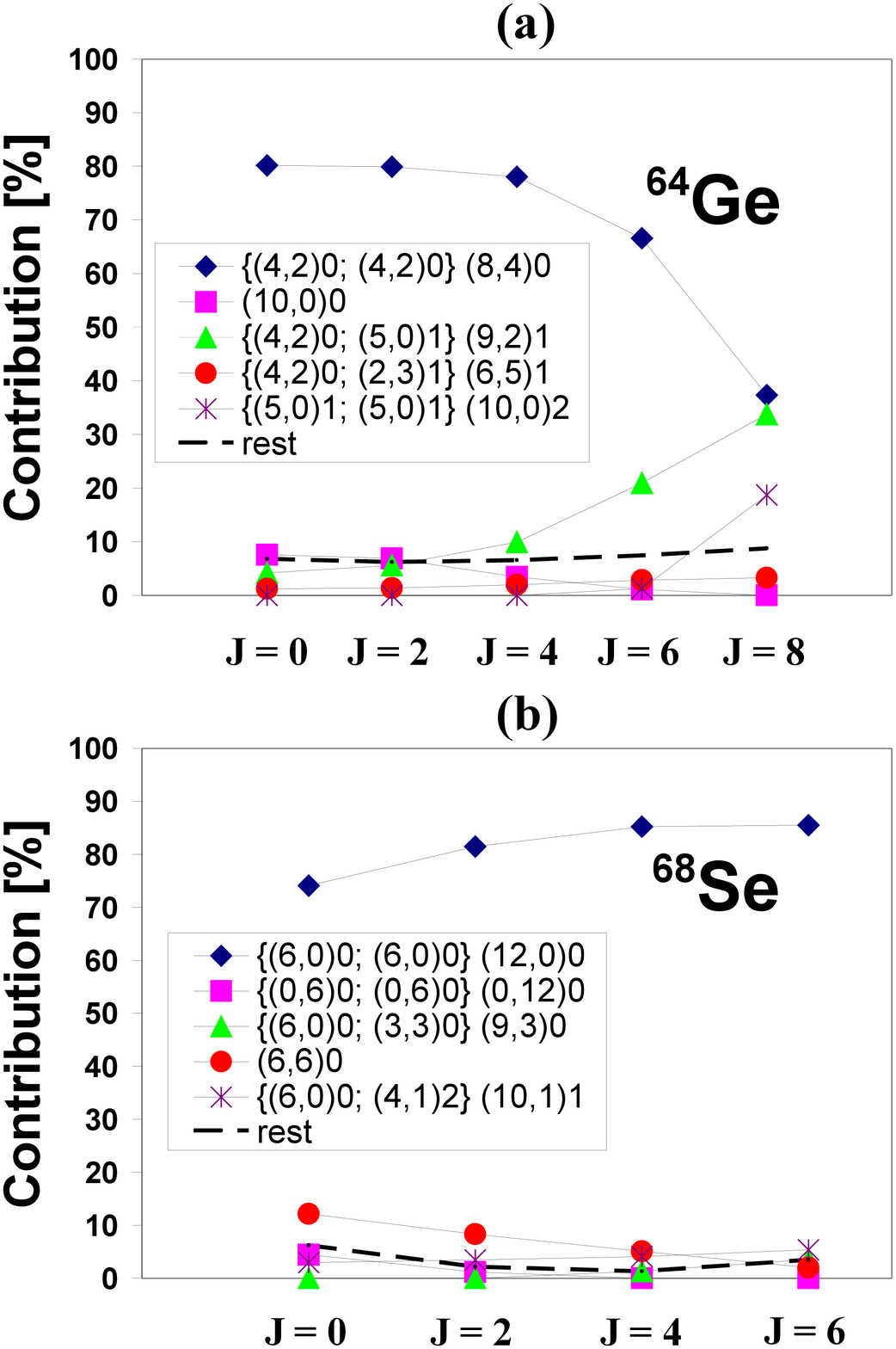} 
\caption{(Color online) Wave-function decomposition of the
calculated extended SU(3) eigenstates in the g.s. band of
(a) $^{64}$Ge and (b) $^{68}$Se. The leading irreps from
the dominant type configurations are listed explicitly while
the effect of those with less than 3\% contribution
for any state as well as from configurations with two and
four particles in the unique space is represented with a
dashed line.}
\label{Figure_wf_content} 
\end{figure}
\begin{table*}[h,t]
\caption{Model-space dimensions for the G-matrix calculations in
the full (restricted) $pf_{5/2}g_{9/2}$ space for $^{64}$Ge ($^{68}$Se)
as well as for the complete $pf_{5/2}$ spaces and for the extended SU(3)
shell model with the irreps listed in Table \ref{TABLE_irreps}.
(The entry marked with a $*$ is smaller by about a factor of two due
to our taking advantage of  time-reversal symmetry, which we did not
invoke in the other cases as machine storage for them was not an issue.)
\label{TABLE_spaces}}
\begin{ruledtabular}
\begin{tabular}{cccccc}
$$ & $$ & $$ & $J$ & $$ & $$ \\
\cline{2-6}
 $$ & $0$ & $2$ & $4$ & $6$ & $8$ \\
\hline
\multicolumn{6}{c}{$^{64}$Ge} \\
  $pf_{5/2}g_{9/2}$  & $1,831,531$ & $1,728,929$ & $1,454,930$ & $1,090,581$ & $724,318$ \\
  $pf_{5/2}$  & $28,503$ & $24,246$ & $14,760$ & $6,183$ & $1,638$ \\
  Ext. $SU(3)$  & $322$ & $1,421$ & $2,098$ & $2,225$ & $2,208$ \\
\multicolumn{6}{c}{$^{68}$Se} \\
  $pf_{5/2}g_{9/2}$  & $1,929,014^{*}$ & $3,611,680$ & $2,973,404$ & $2,138,391$ & $$ \\
  $pf_{5/2}$  & $93,710$ & $81,122$ & $52,175$ & $37,086$ & $$ \\
  Ext. $SU(3)$  & $397$ & $1,765$ & $2,640$ & $3,115$ & $$ \\
\end{tabular}
\end{ruledtabular}
\end{table*}   

The effect of adding the proton-neutron pairing and the pair-scattering
terms in the Hamiltonian is illustrated in Fig. \ref{Figure_energies_pairing}.
For our choice of model space (Table \ref{TABLE_irreps}), the results for
the g.s. bands demonstrate comparable size effects from both interactions,
especially for $^{64}$Ge, with the only clear exception being the
$4_{1}^{+}$ and $6_{1}^{+}$ states in $^{68}$Se. When a smaller number of
irreps is included in the calculation, the pn-pairing interaction has a much
bigger impact, an effect mainly visible for higher-spin states, while
the role of the pair-scattering terms is strongly diminished.  

Electromagnetic transition strengths are normally calculated  
with the E2 transition operator of the form \cite{T(E2)_def,JPD_sym}:
\begin{eqnarray}
T(E2) &\approx& \sqrt{5/16 \pi} A^{1/3} (e_{\pi}\frac{\tilde{\eta_{\pi}}+1}
{\eta_{\pi}}\tilde{Q}_{\pi N}+e_{\pi}Q_{\pi U} \nonumber \\
&+&e_{\nu}\frac{\tilde{\eta_{\nu}}+1}{\eta_{\nu}}\tilde{Q}_{\nu N}+e_{\nu}Q_{\nu U})
\end{eqnarray}
\noindent Instead, in this work we simply used the single dominant component in the
pseudo-SU(3) expansion of the quadrupole operators in the normal-parity space.
The effective charges $e_{\pi}$ and $e_{\nu}$ were taken as $e_{\pi}=1.5$ and $e_{\nu}=0.5$
for the two versions of the realistic interaction and the extended-SU(3) calculations.
The overall agreement between the results for both nuclei (Table
\ref{TABLE_strengths_64Ge} for $^{64}$Ge and Table \ref{TABLE_strengths_68Se} for the g.s. band in $^{68}$Se)
is good, although some recent experimental findings for the $2_{g.s.}^{+}\rightarrow 0_{g.s.}^{+}$ transition strength in
$^{64}$Ge \cite{64Ge_expBE2} seem to be underestimated by approximately a factor of 1.4. The correct 
behavior of the interband transitions is also nicely reproduced. More significant
deviations are observed for the transitions between members of the $K=2^{+}$ band and the 
$(J+1)^{+} \rightarrow J^{+}$ transitions in $^{64}$Ge. These could be attributed to the fact
that some of the states from this band (e.g. $4_{2}^{+}$ and $6_{2}^{+}$) are found to 
be highly mixed with $S=1$ irreps and differ more significantly from the rest thus displaying
a less regular structure pattern throughout the band, for example, to what has been
observed in the same bands of some rare-earth nuclei. It seems that the orbit-orbit interaction is
the part of the Hamiltonian responsible for this feature. Results when only the normal-parity
spaces are included in the calculation (shown in parentheses) reveal a contribution of the
unique-parity sector of only up to 2-3\%. An increase in this number is expected for higher-lying
states or heavier nuclei where the dominant configurations are the ones with an occupied
unique-parity space.    

Finally, let us look at the content of the eigenfunctions for the states in the different bands.
In the g.s. ($K=0^{+}$) band of $^{64}$Ge, one can clearly see the dominance of the leading and
most deformed SU(3) irrep $(8,4)$ (Fig. \ref{Figure_wf_content}(a)) which gradually declines
throughout the band from about 80\% for $J=0_{g.s.}^{+}$ to less than 40\% for $J=8_{1}^{+}$.
Since the spin-orbit interaction is not as strong as in the case of the ds-shell nuclei,
the mixing of irreps is smaller compared to the corresponding normal-SU(3) results for
$^{24}$Mg and $^{28}$Si \cite{Vargas_even}. On the other hand, as mentioned above, the $K=2^{+}$ band
follows a less regular pattern with some of the states being of highly mixed nature. The
$0_{2}^{+}$ state is found to be dominated (85\%) by the $(\lambda,\mu),S=(9,2),1$ irrep. In the case of
$^{68}$Se, the leading irrep $(12,0)$ contributes from 75 to 85\% (Fig. \ref{Figure_wf_content}(b)).
A slight change in the type of Hamiltonian used may help establish the transition from states of prolate
shape dominated by the irrep (12,0) in the g.s. band to ones where the (0,12) irrep prevails. To achieve this
effect we need to add a term proportional to the third order Casimir invariant $C_3$ of SU(3). It
was demostrated earlier \cite{Vargas_even} that this term is capable of adjusting the prolate-oblate
band crossing by driving irreps with $\mu >> \lambda$ lower in energy than those with $\lambda >> \mu$.
The same term can also be used to fix the position of the first excited $0^{+}$ state not assigned yet
experimentally but predicted by our G-matrix calculations to lie at 1.05 MeV.

Although the extended-SU(3) calculations are performed in a model space that involves the 
whole gds shell, the basis is still much smaller in size even compared with the one used for
realistic calculations in the $pf_{5/2}g_{9/2}$ space. This drastic reduction translates into
the use of only hundreds or at most a few thousand basis states (Table \ref{TABLE_spaces}).
For example, the size of the basis used in the extended-SU(3) calculations for
$^{64}$Ge represents only between 0.02\% to 0.3\% of that for unrestricted calculations in the
$pf_{5/2}g_{9/2}$ model space. This means that a space spanned by a set of extended-SU(3)
basis states may be computationally manageable beyond the limit accessible for the modern
full-space shell-model calculations as is the case for the combination of the upper-fp and
the gds shells. 

While some refinements in the model certainly could be done (like trying 
different and more sophisticated types of Hamiltonians, using different
strengths for identical-particle and pn-pairing interactions, etc.)
and the role of the model-space truncation may be further explored, the
results presented in this paper suffice to demonstrate that the SU(3)
scheme in its extended formulation can be a valuable tool for studying
nuclei of the upper-fp-shell region. Its benefits will show even more
prominently in the more general and complicated case, namely, when the
dominant configuration is no longer the one with an empty unique-parity
space and in situations when two or more competing configurations are
closer to one another in energy and as a consequence experience strong mixing. 

\section{Conclusions}
In this paper we extended the usual pseudo-SU(3) shell model for
upper-fp shell nuclei in two ways: firstly by integrating the g$_{9/2}$
level into the dynamics; and secondly by including the entire gds-shell
organized via its SU(3) structure, which we dubbed the extended SU(3)
shell model. While this work only deals with the simplest case in which
one configuration (the one with no particles in the unique-parity space)
dominates all others, it is still possible to appreciate the strength of
this new approach. Specifically, the model offers a richer model space
compared to the previous SU(3) schemes by taking particles from the
unique-parity space explicitly into account. As a result, the current
approach presents an opportunity for a better description of the collectivity
properties of the systems considered by reducing the effective charge
needed in the description of their B(E2) transition strengths. These
results will be even more pronounced for heavier systems where the intruder
space is expected to have higher occupancy. This approach also offers an
opportunity to explore the role of the intruder levels in the dynamics
of the system as in the current study they are treated on the same
footing as the normal-parity orbitals. It is important to underscore that
these advantages are accomplished within a highly truncated and
symmetry-adapted basis, which possibly allows one to reach into otherwise
computationally challenging (if not inaccessible) domains.

The results for the nuclei $^{64}$Ge and $^{68}$Se, presented in this
paper, demonstrate a close reproduction of various results obtained
with a realistic interation. Specifically, many of the states in the
energy spectra and the B(E2) transition strengths are nicely reproduced.
While the results are satisfactory for the states from the g.s. bands,
there still seem to be some need for a more precise description of the
nuclear characteristics related to the properties of the eigenfunctions.
These could be addressed in the future by including some corrections
with the use of more elaborated interactions. Nevertheless, the results
certainly suggest that the extended SU(3) model can be a valuable tool
in studying  properties of nuclei of special interest from this region,
such as those lying close to the proton drip line or/and actively
participating in the processes of nucleosynthesys. They also point to
an excellent opportunity to reveal the role the intruder levels  play in
the dynamics of the system in an exciting and completely new way, namely,
considering their connection to their like-parity partners within the
framework of a severely-truncated symmetry-adapted model space. 

   An extension of the SU(3) shell model for the rare-earth and actinide
nuclei is also underway. It will be able to provide some valuable new
information and a better understanding of the fragmentation and clusterization
phenomena in the B(M1) transition strengths. Also, it will signifficantly
reduce the values of the effective charges used in estimates of the B(E2)
transition strengths. In addition, an expected new emerging structure of
the states in the excited $K=0^{+}$ bands could give an explanation for
the enhanced B(E2) transition strengths to members of the g.s. band.
Finally, the new model provides a powerful means of explanation for the
abundance of low-lying $K=0^{+}$ states found experimentally.

\begin{acknowledgments}
This work was supported by the US National Science Foundation, Grant
Numbers 0140300 and 0500291, and the Southeastern Universities Research 
Association (SURA). We would like to thank Piet Van Isacker for 
providing us with the realistic interactions used.
\end{acknowledgments}
\begin{widetext}
\appendix
\section{\label{appendix:pairing_terms}Monopole pairing and Pair-scattering in the
framework of the extended SU(3) shell model}
The expression for the identical-particle pairing within a shell (and the 
identical-particle pair-scattering between two shells) in terms of SU(3) 
irreducible operators is given in \cite{Bahri_pair}:
\begin{eqnarray}
S^{\sigma \tau \dagger} S^{\sigma \tau^{\prime}} =  
\frac{1}{2} \sum_{
\begin{array}{c}
(\lambda_{1},\mu_{1}) (\lambda_{2},\mu_{2}) \\
\rho (\lambda, \mu)
\end{array}} 
\sum_{l l^{\prime}}(-)^{l-l^{\prime}}
\sqrt{(2l+1)(2l^{\prime}+1)}  
\RedCG{(\eta,0)l}{(\eta,0)l}{(\lambda_{1},\mu_{1})10} 
\RedCG{(0,\eta^{\prime})l^{\prime}}{(0,\eta^{\prime})l^{\prime}}{(\mu_{2},\lambda_{2})10} \nonumber \\
\times \RedCG{(\lambda_{1},\mu_{1})10}{(\mu_{2},\lambda_{2})10}{(\lambda,\mu)10}_{\rho} 
\biggl[ 
\left[ a_{(\eta,0) \frac{1}{2}}^{\dagger} \times a_{(\eta,0) \frac{1}{2}}^{\dagger} \right] ^{(\lambda_{1},\mu_{1})S_{1}=0} \times
\left[ \tilde{a}_{(0,\eta^{\prime}) \frac{1}{2}} \times \tilde{a}_{(0,\eta^{\prime}) \frac{1}{2}} \right]
^{(\mu_{2},\lambda_{2})S_{2}=0} 
\biggr]_{\kappa=1 L=0 \hspace{0.2cm} M_{J}=0}^{\rho (\lambda,\mu), S=0; ~J=0} 
\end{eqnarray}

\noindent where $\sigma=\pi$ or $\nu$ with $\eta = \eta^{\prime}$ ($\eta \neq \eta^{\prime}$) and $\tau=\tau^{\prime}$ 
($\tau \neq \tau^{\prime}$) for the case of pairing (pair-scattering). Here, \RedCG{~}{~}{~} denotes a reduced SU(3)
$\supset$ SO(3) Clebsch-Gordan coefficient and  $\tilde{a}$ is a proper SU(3) 
tensor defined by $\tilde{a}_{(0,\eta)ljm}=(-)^{\eta+j+m}a_{(\eta,0)lj-m}$. 
 
Using the fact that the $9-\lambda \mu$ ($9j$) coefficients connect composite tensors  
corresponding to different coupling schemes of four
SU(3) (SU(2)) tensors, one can derive the  corresponding expressions for
the pn-pairing  (pair-scattering) operators. The final result is given by 
\begin{eqnarray}
S^{\pi\nu, \tau \dagger} S^{\pi\nu, \tau^{\prime}} =  
\frac{1}{4} \sum_{
\begin{array}{c}
(\lambda_{\pi},\mu_{\pi}) (\lambda_{\nu},\mu_{\nu}) \\
\rho^{\prime}(\lambda,\mu)
\end{array}} 
\sum_{\mathcal{L} l l^{\prime} \kappa_{\pi} \kappa_{\nu}}(-)^{l-l^{\prime}}
\sqrt{2 \mathcal{L}+1} \sum_{\mathcal{S}} \sqrt{2 \mathcal{S}+1}  
\RedCG{(\eta,0)l}{(0,\eta^{\prime})l^{\prime}}{(\lambda_{\pi},\mu_{\pi})\kappa_{\pi}
\mathcal{L}} \nonumber \\
\times \RedCG{(\eta,0)l}{(0,\eta^{\prime})l^{\prime}}{(\lambda_{\nu},\mu_{\nu})\kappa_{\nu}\mathcal{L}} 
\RedCG{(\lambda_{\pi},\mu_{\pi})\kappa_{\pi}\mathcal{L}}{(\lambda_{\nu},\mu_{\nu})\kappa_{\nu}\mathcal{L}}
{(\lambda,\mu)10}_{\rho^{\prime}} \nonumber \\
\times \biggl[ 
\left[ a_{(\eta,0) \frac{1}{2}}^{\dagger} \times \tilde{a}_{(0,\eta^{\prime})
\frac{1}{2}} \right] ^{(\lambda_{\pi},\mu_{\pi})\mathcal{S}} \times
\left[ b_{(\eta,0) \frac{1}{2}}^{\dagger} \times \tilde{b}_{(0,\eta^{\prime}) \frac{1}{2}} \right]
^{(\lambda_{\nu},\mu_{\nu})\mathcal{S}} 
\biggr]_{\kappa=1 L=0 \hspace{0.2cm} M_{J}=0}^{\rho (\lambda,\mu), S=0; ~J=0} 
\end{eqnarray}

\noindent where $a^{\dag}(a)$ and $b^{\dag}(b)$ denote proton and neutron
creation (annihilation) operators. Again, $\eta =\eta^{\prime}$
($\eta\neq\eta^{\prime}$) and $\tau = \tau^{\prime}$ ($\tau \neq \tau^{\prime}$) 
for pn-pairing (pn-pair scattering). The labels ($\lambda_{\pi},\mu_{\pi}$) and
($\lambda_{\nu},\mu_{\nu}$) represent
the proton and the neutron SU(3)-coupled irreps and the symbols $\mathcal{L}$
and $\mathcal{S}$ stand for the  orbital and spin angular momentum in the coupled
proton (and neutron) spaces.

\end{widetext}


\begin{thebibliography}{3}
\bibitem{Meyer} 
         M. G. Meyer, Phys. Rev. {\bf 75}, 1969 (1949); {\bf 78}, 16 (1950); 
         {\bf 78}, 22 (1950); 
         O. Haxel, J. H. D. Jensen, and H. E. Suess, {\it ibid.} {\bf 75}, 1766 (1949). 
\bibitem{MonteCarlo} 
         S. E. Koonin, D. J. Dean, and K. Langanke, Phys. Repts. {\bf 577}, 1 (1996).
\bibitem{Elliott}
         J. P. Elliott, Proc. Roy. Soc. London, Ser. {\bf A 245}, 128 (1958); 
         {\bf A 245}, 562 (1958).
\bibitem{JPD}
         R. D. Ratna Raju, J. P. Draayer, and K. T. Hecht, Nucl. Phys. {\bf A202}, 433 (1973).
\bibitem{Escher} 
         J. Escher, J. P. Draayer, and A. Faessler, Nucl. Phys. {\bf A586}, 73 (1995).
\bibitem{Bhatt} 
         K. H. Bhatt, C. W. Nestor, Jr., and S. Raman, Phys. Rev. C {\bf 46}, 164 (1992).
\bibitem{T(E2)_def} 
         K. H. Bhatt, S. Kahane, and S. Raman, Phys. Rev. C {\bf 61}, 034317 (2000).  
\bibitem{Aberg} 
         S. \AA berg, H. Flocard, and W. Nazarewicz, Annu. Rev. Nucl. Part. Sci. {\bf 40}, 469 (1990).
\bibitem{Vargas_even}
         C. Vargas, J. G. Hirsch, and J. P. Draayer, Nucl. Phys. {\bf A690}, 409 (2001). 
\bibitem{Vargas_odd} 
         C. E. Vargas, J. G. Hirsch, and J. P. Draayer, Nucl. Phys. {\bf A697}, 655 (2002).
\bibitem{SU(3)_applications} 
         G. Popa, J. G. Hirsch, and J. P. Draayer, Phys. Rev. C {\bf 62}, 064313 (2000); 
         C. Vargas, J. G. Hirsch, T. Beuschel, and J. P. Draayer, Phys. Rev. C {\bf 61}, 031301(R) (2000).
\bibitem{Schatz}
         H. Schatz et al., Phys. Rep. {\bf 294}, 167 (1998); J. A. Clark et al., Phys. Rev. C {\bf 75}, 
         032801(R) (2007).
\bibitem{Kaneko_68Se}
         K. Kaneko, M. Hasegawa, and T. Mizusaki, Phys. Rev. C {\bf 70}, 051301(R) (2004).
\bibitem{Sun}
         Y. Sun, M. Wiescher, A. Aprahamian, J. Fisker, Nucl. Phys. {\bf A758}, 765c (2005).
\bibitem{Caur}
         E. Caurier, F. Nowacki, A. Poves, and J. Retamosa,  Phys. Rev. Lett. {\bf 77}, 1954 (1996).
\bibitem{Abzouzi}
         A. Abzouzi, E. Caurier, and A. P. Zuker, Phys. Rev. Lett. {\bf 66}, 1134 (1991).
         F. Nowacki, Ph. D. thesis, ULP Strasbourg, 1995 (unpublished). 
\bibitem{Vincent} 
         S. M. Vincent et al., Phys. Lett. B {\bf 437}, 264 (1998).
\bibitem{Van Isacker} 
         P. Van Isacker, O. Juillet, and F. Nowacki, Phys. Rev. Lett. {\bf 82}, 2060 (1999). 
\bibitem{Drumev} 
         K. P. Drumev et al., to be submitted to Nucl. Phys. A.
\bibitem{Whitehead} 
         R.R. Whitehead and A. Watt, J. Phys. G {\bf 4}, 835 (1978); R.R. Whitehead et al., in {\it Theory and
         Application of moment methods in Many-Fermion systems}, edited by B. J. Dalton et al. (Plenum, New York, 1980); 
         E. Caurier, A. Poves, and A. P. Zuker, Phys. Lett. {\bf B252}, 13 (1990); Phys. Rev. Lett. {\bf 74}, 1517 (1995). 
\bibitem{Quasi_SU(3)}
         A. P. Zuker, J. Retamosa, A. Poves, and E. Caurier, Phys. Rev. C {\bf 52}, R1741 (1995); 
         G. Martinez-Pinedo, A. P. Zuker, A. Poves, and E. Caurier, Phys. Rev. C {\bf 55}, 187 (1997); 
         C. Vargas, J. G. Hirsch, P. O. Hess, and J. P. Draayer, Phys. Rev. C {\bf 58}, 1488 (1998).
\bibitem{Ring}
         P. Ring and P. Schuck, \textit{The Nuclear Many-Body Problem} (Springer, Berlin 1979).
\bibitem{Kaneko_64Ge} 
         K. Kaneko, M. Hasegawa, and T. Mizusaki, Phys. Rev. C {\bf 66}, 051306(R) (2002).
\bibitem{Zuker}
         M. Dufour, A. P. Zuker, Phys. Rev. C {\bf 54}, 1641 (1996). 
\bibitem{NNDC}
         National Nuclear Data Center, \\
         http://www.nndc.bnl.gov.
\bibitem{Sahu} 
         K. C. Tripathy and R. Sahu, Int. Journ. Mod. Phys. E {\bf 11}, 531 (2002).
\bibitem{Fischer} 
         S. M. Fischer, D. P. Balamuth, P. A. Hausladen, C. J. Lister, M. P. Carpenter, D. Seweryniak, 
         and J. Schwartz, Phys. Rev. Lett. {\bf 84}, 4064 (2000).
\bibitem{JPD_sym}  
         J. P. Draayer and K. J. Weeks, Ann. Phys. (N.Y.) {\bf 156}, 41 (1984); 
         O. Casta$\tilde{n}$os, J. P. Draayer, and Y. Leschber, {\it ibid.} {\bf 180}, 290 (1987). 
\bibitem{64Ge_expBE2}
         K. Starosta et al., Phys. Rev. Lett. {\bf 99}, 042503 (2007).
\bibitem{Bahri_pair}
         C. Bahri, J. Escher, and J. P. Draayer, Nucl. Phys. {\bf A592}, 171 (1995).
\end{thebibliography}
\end{document}